%% file: coine20.tex
\documentclass[runningheads]{llncs}
\usepackage{graphicx}
\usepackage{colortbl}
\usepackage{booktabs}
\usepackage{amssymb}
\usepackage{longtable}
\usepackage{tabularx}
\usepackage{array}
\usepackage{tabu}
\usepackage{multirow}
\usepackage{siunitx}
\usepackage[caption=false]{subfig}
\usepackage{enumitem}
\usepackage{float}
\usepackage{placeins}
\usepackage{pbox}
\usepackage{makecell}
\usepackage{varioref}
\usepackage{hyperref}
\usepackage{cleveref}
\usepackage[T1]{fontenc}
\usepackage{xcolor}
\usepackage{url}
\usepackage{wrapfig}
\usepackage[title]{appendix}
\usepackage{refcount}
\newcolumntype{R}{>{$}r<{$}}

\usepackage[shortcuts]{extdash}

\begin{document}
\title{Should I tear down this wall? Optimizing social metrics by
evaluating novel actions}
\titlerunning{Should I tear down this wall?}
\author{J\'anos Kram\'ar \and Neil Rabinowitz \and Tom Eccles \and Andrea Tacchetti\\\email{\{janosk,ncr,eccles,atacchet\}@google.com}\\\institute{Deepmind, London, United Kingdom}}

\institute{Deepmind, London, United Kingdom}

\authorrunning{J. Kram\'ar et al.}
\maketitle              %
\begin{abstract}
One of the fundamental challenges of governance is deciding when and how to intervene in multi-agent systems in order to impact group-wide metrics of success. This is particularly challenging when proposed interventions are novel and expensive. For example, one may wish to modify a building's layout to improve the efficiency of its escape route. Evaluating such interventions would generally require access to an elaborate simulator, which must be constructed ad-hoc for each environment, and can be prohibitively costly or inaccurate.
Here we examine a simple alternative: Optimize By Observational Extrapolation (OBOE).
The idea is to use observed behavioural trajectories, without any interventions, to learn predictive models mapping environment states to individual agent outcomes, and then use these to evaluate and select changes. %
We evaluate OBOE in socially complex gridworld environments and consider novel physical interventions that our models were not trained on. We show that neural network models trained to predict agent returns on baseline environments are effective at selecting among the interventions. %
Thus, OBOE can provide guidance for challenging questions like: ``which wall should I tear down in order to minimize the Gini index of this group?''
\end{abstract}
\input{body-conf}
\bibliographystyle{splncs04}
\bibliography{bibliography}

\end{document}

%% file: body-conf.tex
\section{Introduction}

A key interest in multi-agent research is to understand mechanisms that shift group behaviour towards some desirable goals or outcomes such as using institutions, or central agents, that can modify the environment in real time.

For example, suppose there's a traffic intersection that has inconvenienced many road users with its lengthy wait times and poor visibility, and we're advising a municipal government about whether to intervene, e.g. by installing a roundabout. There can be several desiderata, e.g. we may wish to reduce accidents and congestion.

One way to answer the question is to install the roundabout and measure the metrics of interest. This is scientifically simplest but practically prohibitive. Alternatively, one could simulate the change using a detailed traffic model. The model may need to make substantial assumptions that might be broken by unanticipated circumstances, such as weather conditions, or unusual road users like electric scooters, etc. Further, designing and adjusting the simulator is a significant up-front research cost.
Instead, the strategy we propose is to exploit the abundance of observational data (such as traffic patterns across a set of urban environments) and avoid relying on expensive experimental data (in which the intervention is tried, in either reality or simulation). %

Our contribution is a method, which we call OBOE (Optimize By Observational Extrapolation), for building a central agent tasked with altering the physical environment where a group of agents interact, in pursuit of a group-level outcome. We construct this central agent by training a model to predict forward outcomes for each agent (e.g. forward returns) from environment states using purely observational data. Our central agent then uses these models to evaluate candidate interventions in real-time by modifying a snapshot of the environment, and selecting the intervention that produces the best result as estimated by the predictive model, and in terms of the desired outcome metric.

We evaluate our method on two socially complex grid-world environments, a variety of candidate interventions, and consider various model architectures. Our results show that predictive models trained on observational data can identify interventions that lead to desirable outcomes, and are thus well suited to build the central agent we set out to design. In some cases, we find that the resulting OBOE central agent is more effective than a simulation-based central agent that selects an intervention using multiple perfect (but stochastic) simulations of the environment. %

\section{Related work}

Multi-agent reinforcement learning has received considerable attention in recent years, both as a training paradigm to construct powerful individual agents \cite{DBLP:journals/corr/abs-1909-07528,DBLP:journals/nature/SilverHMGSDSAPL16,Balduzzi2018}, and as a tool to study the role and emergence of human pro-social inductive biases and institutions in complex environments \cite{DBLP:conf/nips/HughesLPTDCDZMK18,Leibo2017,Lerer2017,DBLP:conf/atal/PeysakhovichL18}. In this context, the idea of ``opponent shaping'', that is, constructing agents that actively influence the learning and behavior of others through conditional cooperation, or assuming the role of a central mechanism, has emerged as a key challenge to investigate \cite{Bauman2018,Foerster2017,Balduzzi2018,DBLP:conf/atal/MguniJSMCC19,Jaques2019,DBLP:conf/iclr/LetcherFBRW19}. Unlike past work in this area, we investigate how to train such a central mechanism by generalizing entirely from observational data.

The recent advances of multi-agent reinforcement learning, and the challenges posed by agents that interact with other non\-/stationary, adaptive learners has sparked interest in ``machine theory of mind'' \cite{DBLP:conf/icml/RabinowitzPSZEB18}, as well as in the development of sophisticated behavioral models of multi-agent systems, often based on relational architectures \cite{Battaglia2018,DBLP:conf/nips/Hoshen17,DBLP:conf/icml/KipfFWWZ18,DBLP:conf/iclr/TacchettiSMZKRG19}. We build on this work, adapting it for use as a central agent to achieve social goals.

The idea of changing elements of multi-agent environments so as to shepherd the behavior of groups towards outcomes that are desirable for the designer is a well-studied problem in economics. In particular, the sub-field of mechanism design has focused on analyzing the effects of changing payoff structures and reward signals, especially in the context of auction design \cite{Mas1995,DBLP:conf/uwec/Varian95,Conitzer2002}. Our work addresses these questions implicitly in its behavioural model, without reference to rationality assumptions -- however it does not address how learning agents might change their policies in response to the central agent, which is a limitation. %

In recent years there has been work on learning to simulate physical systems \cite{DBLP:journals/corr/abs-2002-09405}, make counterfactual physical predictions \cite{Baradel2020CoPhy}, and learn physical controllers \cite{DBLP:conf/siggraph/GrzeszczukTH98}. Our work builds on this line of thought, targeting the challenges posed by a multiagent setting.

Finally, a number of researchers have investigated the effects of modifying the physical environment so as to promote or dissuade certain group behavior, especially in the context of pedestrian modeling and sheep herding \cite{DBLP:journals/simpra/Martinez-GilLF14,JIANG20181,DBLP:conf/aiide/CowlingG10,bennett2012comparative}. We address this question in a broader context where the behavioural models must be learned rather than hand-coded.

\section{Method and experimental set-up} \label{sec:method}

We introduce OBOE, a new method to construct a central agent that can optimize social metrics by intervening in the physical environment. Importantly, we assume that it is either hard or impossible to simulate the effects of these interventions, and therefore it is desirable to learn to estimate their efficacy based solely on an observational dataset, i.e. from trajectories of agents interacting in the environment, without any intervention whatsoever.

The OBOE method will proceed in 2 stages. In Section~\ref{subsec:obsdata} we describe the dataset; then in Section~\ref{subsec:central} we describe the construction of the central agent. Following those subsections we'll describe the evaluation procedure. In Section~\ref{subsec:cfdata} we describe how we generate a complete evaluation dataset of all possible interventions, and in Section~\ref{subsec:taskfilter} we describe how we use this dataset to identify which tasks are most suitable for evaluating the central agent. %

Note that the training of the central agent in Sections~\ref{subsec:obsdata}--\ref{subsec:central} and its evaluation in Sections~\ref{subsec:cfdata}--\ref{subsec:taskfilter} are separate procedures. Evaluation requires counterfactual data for all possible interventions, which are not generally available in practice. For the purposes of this paper, we provide the counterfactual data (from additional simulations) for estimating the maximum achievable effect by a central agent. It's possible to validate our central agent itself much more economically, by only gathering data according to its prescribed interventions.

In our experiments we simplify the task for the central agent by only considering single interventions at a fixed timestep; this already pushes OBOE into new territory and gives a good indication of its promise, while still making it easy to generate a full evaluation dataset. The same central agents could also be used to intervene at multiple timesteps per episode.

\subsection{Collect observational dataset}\label{subsec:obsdata}

We first need to select environments and collect a dataset of many episodes, with no central agent nor interventions. %
Our environments and players are described in Section~\ref{sec:tasks} and detailed further in Appendix~\ref{appendix:tasks}; in brief, we trained players with multi-agent reinforcement learning \cite{Leibo2017} on some Markov social dilemma games.

\subsection{Construct a central agent}\label{subsec:central} %

We consider a central agent whose objective is to optimise an episode social metric of interest $M$, which is computed per-episode as a function of individual agent' outcomes aggregated across the episode. In our environments these outcomes are the total returns across the episode: for agent $i$, we write this as $R_i=\sum_{t=1}^{T} r_i^t$, where $r_i^t$ is reward. We can then write the metric as $M(R_1,\dots,R_k)$. For example, the collective return is $M(R_1,\dots,R_k)=\sum_i R_i$.
We also need the representation of environment states $s$ to let us construct intervention functions $s'_a(s)$ for each possible intervention $a$ (e.g., build a wall) in the central agent's action space.

\begin{figure}[!b]
\centering
\includegraphics[width=.75\linewidth, keepaspectratio]{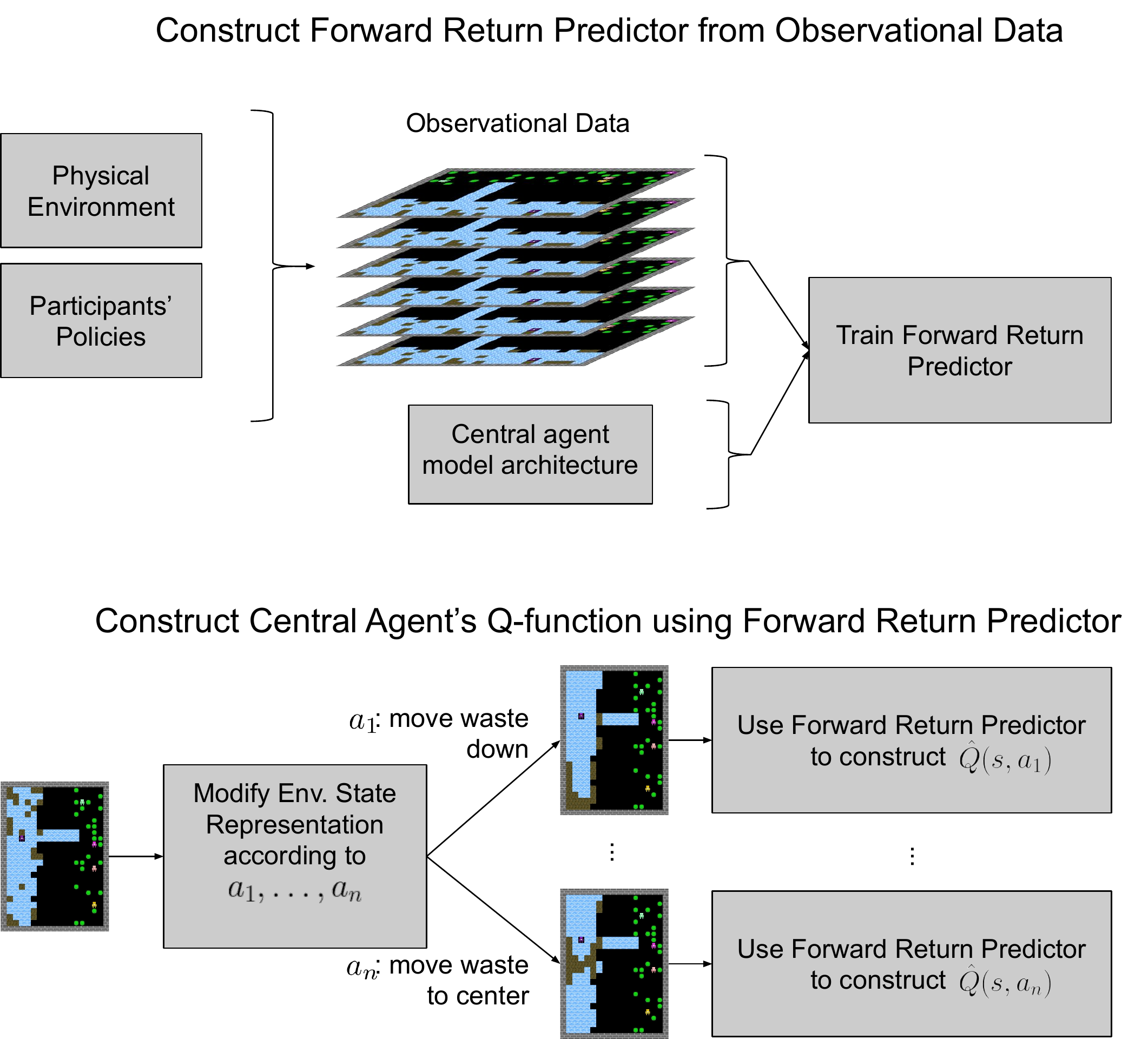}
\caption{An illustration of the OBOE method, with an example ``move waste'' intervention on the Cleanup game.}
\end{figure}

We use the observational dataset from Section~\ref{subsec:obsdata} to train a forward return predictor\footnote{This requires a model; we discuss model choices in Section~\ref{sec:models}.} (or in general, forward aggregated outcome predictor) $\hat R^{>t}(s_t;i)$ (where the environment state $s_t$ includes the timestep $t$, as well as the returns $R_i^{\leq t}(s_t)$ so far for agent $i$). Now, we can combine this with the intervention functions to estimate the central agent's action-value function $Q(s,a)$, i.e. the metric value that would result from the intervention $a$, as $\hat Q(s_t,a)=M(\hat R^{>t}(s'_a(s_t);1)+R_1^{\leq t}(s_t), \dots, \hat R^{>t}(s'_a(s_t);k)+R_k^{\leq t}(s_t))$. We propose a central agent policy that selects actions greedily according to this $\hat Q$.

\subsection{Collect evaluation dataset}\label{subsec:cfdata}
In order to evaluate the central agent, we collect an evaluation dataset. We record new trajectories of game play using, as before, our trained policies for participant agents. Importantly, however, here we do modify the environment state according to all possible interventions, and collect the various outcomes. With this data, we can compare the intervention selected via $\hat Q(s, a)$ to other possible actions, and evaluate the central agent.

\subsection{Task filtering}\label{subsec:taskfilter}
Finally, we use this second dataset to detect intervention families that result in no effect on social outcomes, or which do not require access to the environment states to be evaluated. We remove these tasks from our analysis since they would not be informative. This procedure is outlined in detail in Section~\ref{sec:task_analysis}.

\section{Tasks}\label{sec:tasks}

In order to test the OBOE method, we need suitable tasks, where:
\begin{itemize}
    \item The agents interact nontrivially, so it's not obvious how an intervention will affect social outcomes.
    \item The environment state is fully observed, as in a Markov game.
    \item The observed environment state is easily represented in a structured form, so we can intervene on the structured representation and get predicted intervention effects from the model.
    \item We can simulate the ground truth with the interventions (for evaluation).
\end{itemize}

We use two spatiotemporal social dilemma games which meet the above criteria: Cleanup, and Harvest. \cite{DBLP:conf/nips/HughesLPTDCDZMK18,Jaques2019,DBLP:conf/atal/EcclesHKWL19,DBLP:conf/atal/WangHFCDL19,DBLP:conf/nips/PerolatLZBTG17} In particular, the social outcomes are nontrivial to predict, both because the games pose social dilemmas to the players, and because the players are trained via deep RL. These games allowed us to construct several qualitatively different sorts of interventions, which was another desired criterion. The way the players were trained is detailed in Appendix~\ref{appendix:tasks}. While these games and players were adequate for our purposes, they're not special; OBOE is equally applicable in settings quite unlike them.

\begin{figure}[bt]
\centering
\parbox{0.35\linewidth}{
\centering
\subfloat[before]{
\includegraphics[width=0.45\linewidth, keepaspectratio]{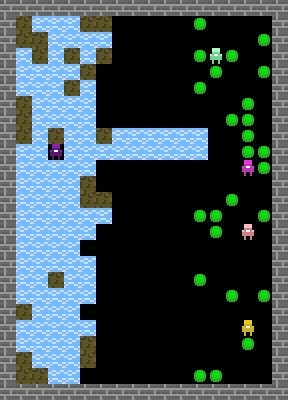}
}\subfloat[after]{
\includegraphics[width=0.45\linewidth, keepaspectratio]{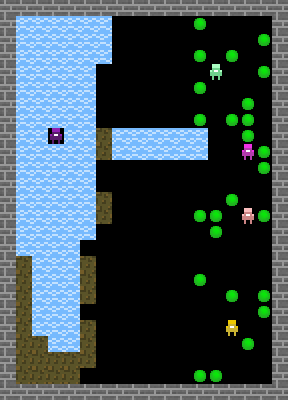}
}
\caption{A ``move waste'' intervention for the Cleanup game.}
\label{fig:cleanup}
}\qquad \parbox{0.55\linewidth}{
\centering
\subfloat[before]{
\includegraphics[width=0.45\linewidth, keepaspectratio]{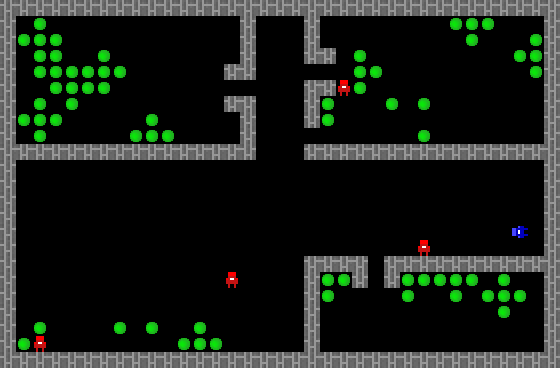}
}\subfloat[after]{
\includegraphics[width=0.45\linewidth, keepaspectratio]{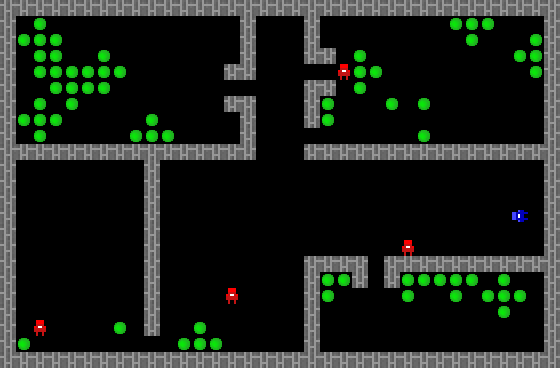}
}
\caption{An ``add wall'' intervention for the Harvest game.}
\label{fig:harvest}
}
\end{figure}

\paragraph{Cleanup:} Cleanup (Figure~\ref{fig:cleanup}) is a public goods game in which 5 players collect apples from a field (right side), which grow only if the players collectively keep the waste accumulating in the aquifer (left side) at a low enough level, by cleaning it up. %
The simple primary objective of players in Cleanup is to collect as many apples as they can. However, the mechanics of the environment place players in a socially complex world: if the waste level is low, it's in each player's individual interest to remain in the field collecting apples. This neglect causes waste to accumulate, which negatively impacts the future apple spawn rate. This intricate dependency poses a dilemma between free-riding (collecting apples) and looking after the public good (cleaning the aquifer), which the players need to navigate.

\paragraph{Interventions:} We consider three types of intervention for a central agent to make partway through a Cleanup game, at time 325 (out of 1000): 1) moving players (to one of 7 predefined locations\footnote{\label{fn:move_locations}(1, 1), (1, 23), (16, 1), (16, 23), (9, 1), (9, 9), and (9, 23) on the $18\times 25$ map.
}, 2) moving waste (falling up, down, left, right, or towards the vertical center, among the aquifer map locations), and 3) moving apples (similarly to moving waste, but among the field locations). A ``move waste'' intervention is illustrated in Figure~\ref{fig:cleanup}.

\paragraph{Harvest:} Harvest (Figure~\ref{fig:harvest}) is a common pool resource game. As in Cleanup, the players collect apples from fields; but in Harvest, the apples only respawn if there are other intact apples nearby. The fields are located in corner rooms of the map, which is procedurally generated; and players can guard their rooms by firing a punishment beam at each other. In Harvest, players face a dilemma between their short- and long-term interests: if they don't distribute their harvesting across space or time, they deplete the apple stock, which may not recover; if they move around too much, other players will collect most apples.

\paragraph{Interventions:} We considered two types of intervention in Harvest: adding walls, and removing walls, both at time 30 (out of 1000). Since the underlying map is variable, we couldn't use a fixed set of interventions lest we tear down a nonexistent wall or build a wall atop a player, so we instead queried the central agent on 15 valid, procedurally generated candidates. An ``add wall'' intervention is illustrated in Figure~\ref{fig:harvest}.

More details about the games are provided in Appendix~\ref{appendix:tasks}.

\subsection{Metrics}

To construct tasks for the central agent, we combine each of these 5 classes of intervention with 4 social outcomes, based on maximizing or minimizing. Two metrics are taken across each episode: the collective return, and the Gini index \cite{gini1912variabilita} (an inequality measure).\footnote{\label{fn:gini}Because the Gini index is nonlinear, strictly speaking the method of Section~\ref{subsec:central} will produce a biased result even if the forward return predictors are unbiased. We disregard this issue.} This gives a total of 20 intervention tasks for the central agent. Table~\ref{tab:interventions} shows the mean metric values among the observational data; equivalently, these give the absolute performance of a ``no-op'' central agent.

\begin{table}[htb]

\begin{tabular}{p{1in}ll}
\toprule
&Cleanup & Harvest \\
\midrule
Maps&standard&procedural (rooms)\\
\midrule[\cmidrulewidth]
\raggedright Baseline mean metrics&
\parbox[t]{.35\textwidth}{\raggedright collective return 1570.2\\Gini index 0.1966}
&
\parbox[t]{.35\textwidth}{\raggedright collective return 584.6\\Gini index 0.3139}\\
\midrule[\cmidrulewidth]
Inter\-ventions&\parbox[t]{.35\textwidth}{\parskip = -\baselineskip\raggedright at $t=325$ of 1000:\begin{itemize}[leftmargin=*]\item move players (each of the 5 players, to 7 fixed locations\footnotemark[\getrefnumber{fn:move_locations}])\item move waste (up, down, left, right, center)\item move apples (similarly)\end{itemize}}&\parbox[t]{.35\textwidth}{\parskip = -\baselineskip\raggedright at $t=30$ of 1000:\begin{itemize}[leftmargin=*]\item add walls (15 random options provided)\item remove walls (15 random options provided)\end{itemize}}\\
\bottomrule\end{tabular}

\caption{A summary of the interventions on the 2 games.}
\label{tab:interventions}
\vspace{-30pt}
\end{table}

\section{Task analysis and filtering}
\label{sec:task_analysis}

In this section we outline a brief analysis of the 20 intervention tasks just described. Our main goal here is to filter out tasks that do not probe the central agent's ability to consider and evaluate modifications to the physical environment in pursuit of an effect on some social outcome. In particular, we seek to select out tasks that fail to meet two intuitive standards: \begin{enumerate}
    \item It should be possible, by modifying the physical environment in the ways prescribed (e.g. by moving agents or adding walls), to affect the social outcome of interest (e.g. to maximize the welfare of our agents). In other words, we wish to prune out those tasks for which even an ``ideal'' central agent cannot help the situation better than by selecting a random action.
    \item We are only interested in tasks where the optimal action depends on the environment state; for example, if we found that for a specific metric, no matter the state of the game, it's always best to move agents to the top left corner, we would filter out this task.
\end{enumerate}

Here we describe the selection process, and the details of how we constructed three baseline central agents: an ``ideal'' one, a ``random'' one and a ``constant action'' one. These are constructed using an evaluation dataset with game episodes for both environments we consider ($4\times 250$ for Cleanup\footnote{Corresponding to the different player populations; see Appendix~\ref{appendix:tasks_cleanup}.}, 5000 for Harvest), in which every possible intervention was tried.

\paragraph{The ``random'' central agent:} We measured the average outcome of all possible interventions in each scenario.

\paragraph{The ``constant action'' central agent:} We measured the performance of each intervention across all recorded episodes, and selected the intervention whose average was best\footnote{This creates a small selection bias.}. We highlight that in Cleanup the interventions are defined across episodes (e.g. ``move waste up''), whereas in Harvest the only consistent constant intervention is ``do nothing'', because the ``add wall'' and ``remove wall'' locations are procedurally generated.

\paragraph{The ``ideal'', or ``CV'' central agent:}
As in OBOE, this central agent was constructed to select the candidate intervention based on its predicted effect on social outcomes. However, instead of extrapolating using a model, we used the \emph{true environment simulator}: that is, we selected the best intervention according to the actual environment trajectories we recorded. Since both games and participants' policies are stochastic, doing this na\"ively results in a selection bias; we eliminated this bias by using cross-validation. For every intervention we considered, we generated 5 different post-intervention completions of the episode by altering a random seed that's used for sampling from the players' stochastic policies. One of the 5 trajectories was used for evaluation, and the other 4 were used to produce an unbiased estimate of the social outcome (by averaging).

\begin{table}[tb]
\centering

\begin{tabular}{rrrrrrrrr}
\toprule
\textbf{Cleanup}\footnotemark[\getrefnumber{fn:apples}]&\multicolumn{4}{c}{move players (36 choices)}&\multicolumn{4}{c}{move waste (6 choices)}%
\\
\cmidrule(l){2-5}\cmidrule(l){6-9}%
&\multicolumn{2}{r}{collective}&\multicolumn{2}{r}{Gini index}&\multicolumn{2}{r}{collective}&\multicolumn{2}{r}{Gini index}%
\\
\cmidrule(l){2-3}\cmidrule(l){4-5}\cmidrule(l){6-7}\cmidrule(l){8-9}%
&min&max&min&max&min&max&min&max%
\\
\midrule
Random&$-4.5$&$-4.5$&0.0026&0.0026&6.8&6.8&0.0001&\cellcolor[HTML]{ea9999}0.0001%
\\
Best constant&$-54.6$&\cellcolor[HTML]{ffe599}11.3&$-0.0023$&0.0295&$-5.2$&\cellcolor[HTML]{ffe599}29.1&\cellcolor[HTML]{ffe599}$-0.0004$&\cellcolor[HTML]{ffe599}0.0015%
\\
\midrule
CV&\cellcolor[HTML]{b6d7a8}$-71.8$&12.6&\cellcolor[HTML]{b6d7a8}$-0.0039$&\cellcolor[HTML]{b6d7a8}0.0481&\cellcolor[HTML]{b6d7a8}$-14.1$&23.6&$-0.0004$&0.0005%
\\
\midrule
\textbf{Harvest}&\multicolumn{4}{c}{add walls (16 choices)}&\multicolumn{4}{c}{remove walls (16 choices)}\\
\cmidrule(l){2-5}\cmidrule(l){6-9}
&\multicolumn{2}{r}{collective}&\multicolumn{2}{r}{Gini index}&\multicolumn{2}{r}{collective}&\multicolumn{2}{r}{Gini index}\\
\cmidrule(l){2-3}\cmidrule(l){4-5}\cmidrule(l){6-7}\cmidrule(l){8-9}
&min&max&min&max&min&max&min&max\\
\midrule
Random&$-1.5$&$-1.5$&0.0152&0.0152&12.9&12.9&$-0.0007$&$-0.0007$\\
Null&0&0&0&0&\cellcolor[HTML]{ffe599}0&0&0&0\\
\midrule
CV&\cellcolor[HTML]{b6d7a8}$-99.1$&\cellcolor[HTML]{b6d7a8}71.2&\cellcolor[HTML]{b6d7a8}$-0.0219$&\cellcolor[HTML]{b6d7a8}0.0993&3.1&\cellcolor[HTML]{b6d7a8}37.8&\cellcolor[HTML]{b6d7a8}$-0.0093$&\cellcolor[HTML]{b6d7a8}0.0092\\
\bottomrule
\end{tabular}
\caption{Filtering intervention tasks. Each intervention task is determined by a game, a class of interventions (which always includes the null intervention), a social metric, and a direction to optimize the metric. For each task, we compare the mean effect of the CV central agent with two baselines. Where we have statistically significant evidence that CV outperforms both (at level $\alpha=0.05$), we highlight the CV result; where we don't, we highlight the baseline(s) it didn't significantly outperform, with red for the random baseline (suggesting the intervention may have no effect at all) and yellow for the constant baseline (suggesting it may not be possible to do better than selecting the same intervention every time).}
\label{tab:cv_effects}
\vspace{-10pt}
\end{table}

The CV central agent provides an interesting benchmark. It reflects the performance we'd expect if we had access to a perfect simulator of the interventions, and a reasonable (if na\"ive) budget to run such simulations. This allows us to select out tasks for which even an ``ideal'' agent cannot do better than random, or better than a ``constant action'' agent, and to put other central agents' performance into context. In Table~\ref{tab:cv_effects} we present the achieved effect for each task\footnote{\label{fn:apples}The ``move apples'' tasks in Cleanup were excluded because even the CV central agent could not produce any significant effect.} (relative to choosing no intervention) of selecting using the CV central agent, versus these other baselines. To conduct the task filtering, the table indicates for each task whether the CV central agent outperformed the baselines with statistical significance (at level $\alpha=0.05$).

\section{Models}
\label{sec:models}
As described in Section~\ref{subsec:central}, we construct the central agent using models that predict the forward returns of each player, given a time step observation of the games. We use three architectures: two ubiquitous ones (a convolutional neural network (CNN) and a multi-layer perceptron (MLP)), and a Relational Forward Model (RFM) \cite{DBLP:conf/iclr/TacchettiSMZKRG19}, chosen because it captures multi-agent interactions well and its structured input representation makes it easy to construct interventions. Details about model hyperparameters and training are provided in Appendix~\ref{appendix:model}. %
The three architectures call for slightly different input representations. %

\begin{itemize}

\item For the RFM, we use a graph representation whose nodes are the individual agents as well as the potentially active map locations (all locations for Harvest, apple and waste locations for Cleanup). The node features are: x and y position, a one-hot for the non-agent content of the map location (for Cleanup: clean water, dirty water, empty field, apple; for Harvest: empty field, apple, wall, empty space), the player's identity (for Cleanup, i.e. prosocial or antisocial; see Appendix~\ref{appendix:tasks_cleanup}), the last action and reward, and the player orientation. The graph contains edges from every node to every agent node, and no input edge features are provided. Further, there is a single global feature: the timestep. All features are whitened to have mean zero and standard deviation 1.

\item For the MLP, we use an input representation that contains all of the nodes' features, and the global feature, encoded as a concatenated vector.

\item For the CNN, we provide separate channels for the agent node features, the non-agent node features, the global feature (timestep), and a channel that indicates which agent to predict forward return for.

\end{itemize}

\section{Results}

In Section~\ref{subsec:res_tasks} we describe the outcome of the task filtering we laid out in Section~\ref{sec:task_analysis}. In Section~\ref{subsec:res_models} we briefly compare the models in terms of their predictive performance on the validation subset of the observational data. In Section~\ref{subsec:res_agents} we'll combine these tasks and models, establishing the resulting OBOE central agents' success on the filtered tasks.

It's useful to note that while the task filtering is useful here (because we don't know a priori what to expect from these interventions), we are only using the CV agent as a source of additional validation. The CV agent allows us to measure performance relative to an estimated optimum, but measuring the OBOE agents' performance in absolute terms would only require data for their selected interventions, rather than for all candidate interventions (as CV does).

\subsection{Tasks}\label{subsec:res_tasks}

Before turning our attention to the trained models and resulting central agents, we review Table~\ref{tab:cv_effects}, in which we probe tasks by comparing the CV central agent with na\"ive baseline agents. On one of the tasks (5 including the ``move apple'' interventions\footnotemark[\getrefnumber{fn:apples}]), the CV agent doesn't significantly beat a random baseline, so it may not be possible to achieve the desired social outcome using the available interventions. For an additional 4 tasks, there was a constant-intervention baseline that the CV agent couldn't significantly beat. This leaves 11 tasks (4 in Cleanup, and 7 of the 8 in Harvest) where we have evidence that the task meets the standards laid out in Section \ref{sec:task_analysis}; we call these the \textit{significant tasks}.

\subsection{Models}\label{subsec:res_models}

\begin{wraptable}[10]{r}{0pt}
\begin{tabular}{llll}
\toprule &MLP&RFM&CNN\\\midrule Cleanup&775&757&748\\Harvest&1485&1291&1205\\\bottomrule
\end{tabular}
\caption{Best validation loss for each architecture in Section~\ref{sec:models}.}
\label{tab:v_loss}
\end{wraptable}

Following, we focus on the performance of the central agent design we propose. We trained models as described in Section~\ref{sec:models}, selecting the ones for each architecture that achieve lowest validation MSE on forward returns; the validation losses are displayed in Table~\ref{tab:v_loss}.

All models achieved comparable validation loss on the no-intervention datasets. For each dataset, the convolutional model achieved lowest loss, followed by the relational model.

\subsection{Central agents}\label{subsec:res_agents}

\begin{wrapfigure}{r}{0pt} %

\includegraphics[width=0.5\linewidth, keepaspectratio]{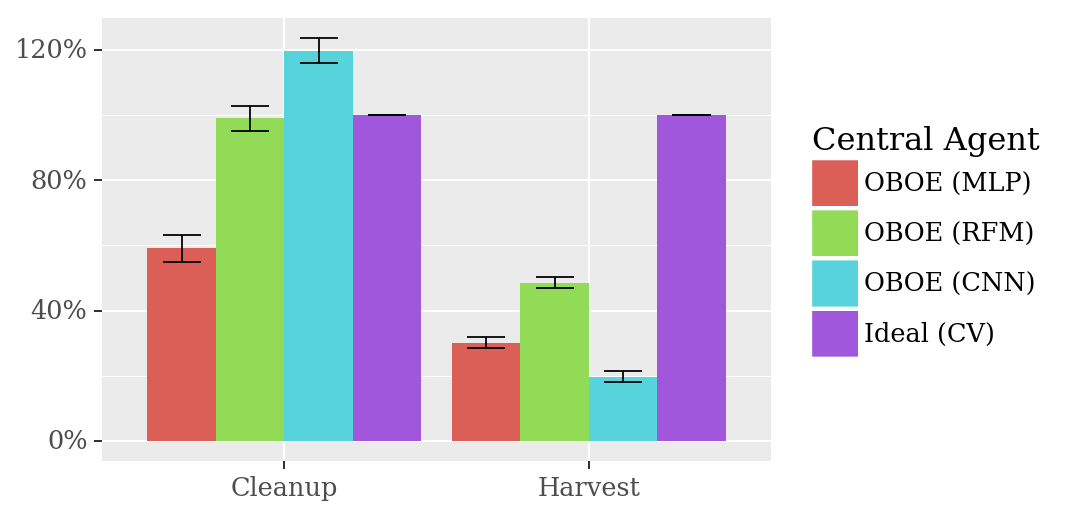}
\caption{Average percent effectiveness relative to CV, across significant tasks. The error bars indicate standard error.}
\label{fig:effects}
\vspace{-\baselineskip}
\end{wrapfigure}

Finally, we evaluate the OBOE central agents constructed using these models, on each of the significant tasks. In order to normalize the performance across the tasks, we compute a normalized score as: $$\mathit{effectiveness}=\frac{\bar M_{\mathit{CA}} - \bar M_{\mathit{random}}}{\bar M_{\mathit{CV}} - \bar M_{\mathit{random}}}~,$$ where $\bar M_{\mathit{CA}}$ is the sample mean of the metric $M$ achieved by a central agent $\mathit{CA}$ for a specific task. Figure~\ref{fig:effects} shows the mean cross-task effectiveness of each model; Table~\ref{tab:effects} shows a full breakdown of the mean effects.

Note that this is a very limited set of games and settings, so we can't generalise any conclusions from comparing the model architectures. Instead the main comparison of interest is between CV (which uses simulations to make an unbiased estimate of each intervention's effect) and the OBOE central agents (which use a learned model to extrapolate from observational data); the striking result is that the OBOE agents are able to do 56\% as well on average as the CV central agent, even though the former are being applied outside their training distributions, and the latter has access to targeted ground-truth simulations.

\begin{table}[!tb]
  \begin{minipage}{\linewidth}
\scalebox{.95}{
  \begin{tabular}{
  llllRRRRR
  }
    \toprule
    &&&&\multicolumn{5}{c}{Mean intervention effect}\\
    \cmidrule(){5-9}
    \multicolumn{2}{l}{Intervention Class}&&&&&\multicolumn{3}{c}{OBOE}\\
    \cmidrule(){1-2}\cmidrule(l){7-9}
    Game&Type&\multicolumn{2}{c}{Task}&\text{Random}&\text{CV}&\text{MLP}&\text{RFM}&\text{CNN}\\
    \midrule
    Cleanup & move players & min & collective return & -4.5 & -71.8 & -60.6 & \mathbf{-96.7} & -95.0\\
    && min & Gini index & 0.0026 & -0.0039 & -0.0002 & -0.0034 & \mathbf{-0.0050}\\
    && max & Gini index & 0.0026 & 0.0481 & 0.0198 & 0.0464 & \mathbf{0.0485} \\
    & move waste & min & collective return & 6.8 & -14.1 & -8.5 & -7.8 & \mathbf{-19.8}
    \\
    Harvest & add wall & min & collective return & -1.5 & -99.1 & -12.1 & -26.0 & \mathbf{-26.3} \\
    && max & collective return & -1.5 & 71.2 & 6.5 & \mathbf{13.4} & 0.0 \\
    && min & Gini index & 0.0152 & -0.0219 & 0.0156 & \mathbf{0.0028} & 0.0185 \\
    && max & Gini index & 0.0152 & 0.0993 & 0.0240 & \mathbf{0.0433} & 0.0262 \\
    & remove wall & max & collective return & 12.9 & 37.8 & 31.2 & \mathbf{44.4} & 42.3 \\
    && min & Gini index& -0.0007 & -0.0093 & \mathbf{-0.0138} & -0.0054 & -0.0032 \\
    && max & Gini index& -0.0007 & 0.0092 & -0.0025 & \mathbf{0.0041} & -0.0036 \\
  \bottomrule
\end{tabular}
}
\end{minipage}
\caption{Mean performance of the OBOE central agents on each task, with the random baseline and CV benchmark. The RFM OBOE agent always outperformed the random baseline -- the other OBOE agents were worse than random on 2 tasks each.}
\label{tab:effects}
\vspace{-15pt}
\end{table}

\section{Discussion and future work}

We introduced OBOE, a novel, widely applicable method for constructing a central agent that selects modifications to a physical environment with multiple interacting agents, to maximize social metrics. In environments where it's prohibitively expensive to test out modifications, or run high-quality simulations of them, OBOE learns from observational data (i.e., recordings of the environment in which no intervention has been made), and constructs a state-action $Q$-function that predicts the effects of physical interventions on the social metric of interest.

We validated OBOE on a range of tasks consisting of different social metrics and games, using different candidate interventions and predictive architectures. Averaging across all of these, OBOE was 56\% as effective as an estimated ideal central agent that was able to perfectly test each modification multiple times.

Notably, our constructed $Q$-function is a causal effect estimate based on correlations in the training set. It's worth discussing why this can work. The main principle is: if the environment (including players) is Markov then causal effects on future outcomes are equal to conditional expectations of those outcomes on the intervention timestep, and if the interventions are within the training distribution then a well-trained model will have low prediction error on them.

In our experiments we relied on two environments, using four families of candidate interventions which were effective in shifting the metrics. The environments were approximately Markov (with the only non-stationarity being the players' recurrent neural network states). However, the interventions were outside the training distribution to varying degrees (e.g. there are no free-floating walls in the training distribution), so the models needed to extrapolate using their inductive biases.

It is worth briefly discussing how actions picked using our model can sometimes lead to a better outcome than actions selected using cross-validation (i.e. constructed using the true simulator as a model). In the high-variance settings we consider, where the inherent stochasticity of individual agents' policies has large effects on social outcomes, a model trained on large and diverse datasets might be better than unbiased simulators that only consider a few realizations.

While we presented OBOE as a replacement for a simulator, it could also be used in conjunction with a simulator that's accurate but expensive to run or design; OBOE could identify promising interventions quickly using a neural network forward pass, permitting better use of the simulation budget. The advantage would be increased if the environment had substantial stochasticity or if there was a larger action space to search through, requiring more simulations.

In the broader context of multi-agent research and simulation, OBOE constitutes an important tool to evaluate the effects planned modifications to the environment might have on complex social metrics.

{\small

\begin{subappendices}

\renewcommand{\thesection}{\Alph{section}}

\section{Model details}\label{appendix:model}
Here we describe hyperparameters and training for the models in Section~\ref{sec:method}.

The RFM is constituted of an ``edge block'' ReLU MLP that computes a representation for each edge given the global and adjacent node features, feeding into a ``node block'' ReLU MLP that computes a representation for each node given the existing node features, global features, and incoming edge features. In the case of Cleanup, we used batch size 640; edge MLP layer sizes 64, 32, 32; and node MLP layer sizes 64, 32, 32, 1. In the case of Harvest, we used batch size 160; edge MLP layer sizes $5\times128$; and node MLP layer sizes $5\times128$, 1.\footnote{The final 1 indicates that there is only one scalar prediction for each node; predictions for non-player nodes are ignored.}

For the MLP models, we again used batch size 640 for Cleanup and 160 for Harvest. The layer sizes for Cleanup were 64, 32, 32, 5, while for Harvest they were 128, 128, 128, 128, 5. (The final 5 corresponds to the 5 players.)

For the CNN models, we used batch size 160 and kernel sizes $3\times3$. For Cleanup, we used 8 layers, with channel counts $2\times 64$, $3\times 128$, $3\times 256$, with strides 1, 1, $2\times(2, 1, 1)$. For Harvest, we used 5 layers, with channel counts $2\times 64$, $3\times 128$, with strides 1, 1, 2, 1, 1. Each layer was followed by a batch normalizer and a ReLU activation, and the final layer was followed by an average pooling layer and a linear layer.

For learning the model parameters we used an RMSProp \cite{tieleman2012lecture} optimizer with learning rate $10^{-4}$, for $2.5\cdot 10^6$ steps, or until convergence. Approximately $2\cdot 10^5$ episodes of data were in the training set for each game; the validation set for each was 1000 episodes. The optimization criterion in every case was the mean squared error relative to the true forward returns of the players in the given episodes.

\section{Tasks}\label{appendix:tasks}

\subsection{Cleanup: game and agents}\label{appendix:tasks_cleanup}

Cleanup is a played on an $18\times 25$ map, in which 5 players collect apples (each worth $+1$ reward) from a field. The field is sustained by a connected aquifer on which waste appears at a constant rate. The apple spawn rate falls linearly with the amount of waste in the aquifer, up to a saturation point where apples do not spawn. At the start of each game episode, no apples are present, and the waste level is just beyond saturation.

Each game episode runs for 1000 time steps. At each timestep, players observe a $15\times 15$ RGB window centered at their position and orientation. Agents can move around or fire a ``fining beam'' or ``cleanup beam'',  both of which have limited range. The fining beam gives $-50$ reward to a player that's within range if there is one, and $-1$ reward to the player using it. The cleanup beam removes waste within its range.

Standard RL algorithms for this task don't find a policy that responds effectively to the dilemma between collecting apples and tending to the public good\cite{DBLP:conf/nips/HughesLPTDCDZMK18}. Since this particular issue is outside the scope of this work, we side-stepped it by letting some agents be intrinsically motivated to clean the aquifer. Specifically, we trained 2 policies with additional reward per unit of waste cleaned: a ``prosocial'' policy that would receive $+1$, and an ``antisocial'' policy that would receive $-1$. These were co-trained using episodes in which players were randomly assigned them in a 1:4, 2:3, 3:2, or 4:1 proportion, using A2C \cite{DBLP:conf/icml/EspeholtSMSMWDF18}. After convergence, these populations behave in a sensible way in the environment and, as a group, collect close to the optimal number of apples in each episode.

\subsection{Harvest: game, procedural generation, and agents}

Harvest is a common pool resource game on a $35\times 23$ map. In Harvest, the apple spawn rate in each field location depends on the number of nearby apples, and falls to zero once there are no apples within a certain radius. Each player's observations and actions are as in Cleanup, except there is no cleanup beam, the fining beam gives a reward of $-30$ to the target agent, and agents are not visually distinguishable (i.e. all agents appear red to other agents, and blue to themselves).

The maps were procedurally generated to have corner rooms of varying size enclosed by walls, sometimes perforated or absent. The exact procedure is as follows. Each corner of the map has a room, randomly assigned height 6 or 9 and width 12 or 15. For each room, we generate walls; 20\% of the time, the wall will be ``perforated'' by letting each wall location remain as a hole with 10\% probability. Next we add an entrance corridor to each room, as in Figure~\ref{fig:harvest}; the location of the entrance must be a wall location that isn't in a corner or adjacent to a hole. (If no such locations exist, we don't add an entrance.) We then generate apple locations: we randomly choose 6\% of eligible locations (rounded down), and add apples at those locations plus their immediate neighbours. All in-room locations are eligible apart from walls, holes, and entrance corridors. Finally, with 30\% probability we randomly sample 1 or 2 (depending on whether height is 6 or 9) eligible locations as player spawn locations, and remove all of the walls. After this has been done for each room, we adjust the number of player spawn locations to 5, by either subsampling, or sampling from the non-room map locations.

In all other respects, the Cleanup and Harvest games are as published at \cite{Vinitsky}. %

In Harvest, we considered two types of intervention: adding walls, and removing walls, at time 30. As discussed in Section~\ref{sec:tasks}, these were procedurally generated to ensure they are valid. This was done by starting from a random suitable map location (wall for removing, non-wall non-player for adding), randomly choosing whether to extend horizontally or vertically, and then randomly choosing how far to extend in that direction. For removing, the length to extend in each direction was chosen uniformly between 0 and the maximum reachable amount of wall. For adding, the maximum length was broken into 3 partitions using a Dirichlet-multinomial distribution with parameters $(\frac{1}{2}, 2, \frac{1}{2})$, and the middle one was used as the wall segment.

As with Cleanup, standard RL algorithms can't find a sustainable policy for Harvest \cite{DBLP:conf/nips/HughesLPTDCDZMK18} so we modified the environment slightly: in our agent training, if a player moves to collect an apple with $k<3$ apples within distance 2, collection would fail with probability $2^{-k}$.

\end{subappendices}}